\documentclass[aps,prl,twocolumn,groupedaddress]{revtex4}
\usepackage[dvips]{graphics}

\begin{document}


\title{Rough-conduit flows and the existence of fully developed turbulence}


\author{G.\ Gioia}
\author{Pinaki Chakraborty}
\affiliation{Department of Theoretical and Applied Mechanics, University of Illinois, Urbana, IL 61801}
\author{Fabi\'an A.\ Bombardelli}
\affiliation{Department of Civil and Environmental Engineering, University of California, Davis, CA 95616}


\begin{abstract}
It is widely believed that at high Reynolds
number (Re) all turbulent flows approach a
state of ``fully developed turbulence"
defined by a unique, Re-independent statistics of
the velocity fluctuations. Yet direct
measurements of the velocity fluctuations have
failed to yield clear-cut empirical evidence of the existence
of fully developed turbulence. Here we relate the
friction coefficient ($f$) of rough-conduit flows
to the statistics of the velocity
fluctuations. In light of
experimental measurements of $f$, our results
yield unequivocal evidence of the existence of
fully developed turbulence.
\end{abstract}


\maketitle


 Turbulence is the unrest that 
 spontaneously takes over a streamline
  flow when the flow is 
 made sufficiently fast. 
 A conspicuous manifestation of this unrest,
 and one that lends itself readily
 to theoretical analysis, is the advent of
 fluctuations in the velocity field of the flow. 
 In a classic paper
 \cite{k41}, Kolmog\'orov 
 made a few plausible assumptions to show 
 that in flows of high mean velocity (or high Re)
 the statistics of the velocity fluctuations
 should become asymptotically invariant to further 
 increases in the mean velocity.
 The universal turbulent state defined by
 the limiting statistics of the 
 velocity fluctuations was termed 
 ``fully developed turbulence,'' and its 
 existence has become a widely held belief.
 Support for this belief 
 has been sought in direct, hot-wire
 measurements of
  the velocity fluctuations; yet, as recent 
 research has shown \cite{po,other,bg,sree},
 the results remain inconclusive. 
In this Letter, we seek to prove the existence
 of fully developed turbulence 
 by harnessing empirical evidence 
 other than the direct measurement of
  the velocity fluctuations. We start with 
 an outline of the intricacies
of the problem. 
 
 Kolmog\'orov studied the 
 statistics of the velocity 
fluctuation at the lengthscale $l$, $u_l$. He used
 dimensional analysis to show that the structure 
 function $\overline{(u_l)^2}$ (i.e., the mean value 
 of $(u_l)^2$) must take the form \cite{k41}
\begin{equation}
  \overline{(u_l)^2} = P[{\rm Re},l/L]\,(\varepsilon l)^{2/3},
\end{equation}
where $P$ is a dimensionless 
 function of the dimensionless variables Re and $l/L$, 
 ${\rm Re} \gg 1$ is a Reynolds number of the flow, $L$ is the 
 largest lengthscale in the flow, $\varepsilon$ is the mean 
  value of the rate of 
 energy dissipation per unit mass, and $l$ is confined to the 
 inertial range, $L \gg l \gg \eta$, where $\eta$ is the viscous 
 (dissipation) lengthscale \cite{orders}.
  Given that ${\rm Re} \gg 1$ and 
 $l/L \ll 1$, it is natural to identify a plausible asymptotic 
 scenario for ${\rm Re} \rightarrow \infty$ and $l/L \rightarrow 0$. 
 To that end, Kolmog\'orov assumed complete similarity with 
 respect to Re and $l/L$, or
 $\lim_{{\rm Re} \rightarrow \infty} \lim_{l/L \rightarrow 0}P=p$, 
 where $p>0$ is a constant prefactor \cite{ba}.
Under this assumption, the leading term in the asymptotic
 expansion of $\overline{(u_l)^2}$  is
$\overline{(u_l)^2}_{\rm lt}= p\, (\varepsilon l)^{2/3}$
  (where the subscript ``lt'' stands for {\it leading term\/}),
  which is {\it independent
 of\/} Re, and may therefore 
be said to define a universal limiting state
 for ${\rm Re} \gg 1$---the state of 
 fully developed turbulence \cite{fd}. 
 Shortly after the publication of Kolmog\'orov's paper 
 in 1941, it was objected that the asymptotic scenario 
 proposed by Kolmog\'orov, and customarily known as K41, 
 could not account for intermittency---a phenomenon whereby 
 the rate of energy dissipation per unit mass 
 fluctuates around its mean value, $\varepsilon$ \cite{frisch}. 
To account for intermittency, Kolmog\'orov himself
 argued that the similarity with respect to $l/L$ might 
 be incomplete, and went on to assume the simplest 
 type of incomplete similarity with respect to $l/L$ \cite{ba}.
 Under this assumption, the leading term of
  $\overline{(u_l)^2}$ is 
  $\overline{(u_l)^2}_{\rm lt}=p\,(\varepsilon l)^{2/3} (l/L)^{\alpha}$, 
where $\alpha$ is a constant intermittency exponent \cite{k62}. 
 This alternative asymptotic scenario, known as K62,
 has led to a vast body of research on
  intermittency \cite{frisch}. 
 Nevertheless, in K62 $\overline{(u_l)^2}_{\rm lt}$ 
 remains independent of ${\rm Re}$, 
 as was the case in K41, and therefore continues to define 
 a universal limiting state for ${\rm Re} \gg 1$.
 In this sense, K62 represents only a minor departure 
 from K41. More recently, a major departure from K41
 has been suggested 
 on the basis 
 of new experimental results \cite{po,other}. 
 These results indicate that even at very high Re 
the prefactor $p$ is not constant, 
 but subject to a discernible dependence 
 on Re---a dependence
  that is marked 
  enough as 
  to cast doubts on
  the existence of fully developed turbulence.
  To analyze these results, Barenblatt and Goldenfeld
  \cite{bg} argued that there might be no similarity
 with respect to Re.
 Further, they argued that the form of the 
 structure function at high Re should be invariant
 under a natural redefinition of Re \cite{natu}, and showed that for this 
 {\it principle of asymptotic covariance\/} to hold 
 the structure function (and therefore 
 $p$) 
 must depend on Re only through $\ln {\rm Re}$.
 Last, they wrote  $p(\ln{\rm Re})=p_0+p_1 \delta + {\rm o}(\delta)$,
 where $\delta\equiv 1/|\ln{\rm Re}|\ll 1$ \cite{inter}. Under these 
 conditions, 
 \begin{equation} \label{asym}
\overline{(u_l)^2} = \left(p_0 +{{p_1}\over{\ln{\rm Re}}}\right) 
 (\varepsilon l)^{2/3} (l/L)^\alpha + 
 {\rm o}\left({{1}\over{\ln{\rm Re}}}\right),
 \end{equation}
and the existence of a universal limiting state depends on 
 the value of $p_0\ge 0$. If $p_0> 0$, 
 then $\overline{(u_l)^2}_{\rm lt}=
 p_0\, (\varepsilon l)^{2/3} (l/L)^{\alpha}$, 
 which is independent of Re,
 and a universal limiting state exists, identical
 with that predicted by K41 (if $\alpha=0$) 
 or K62 (if $\alpha\neq 0$) \cite{uni}. On the other hand, if
 $p_0=0$, then $\overline{(u_l)^2}_{\rm lt}= 
 p_1\,(\varepsilon l)^{2/3} (l/L)^{\alpha}/\ln{\rm Re}$,
 which depends on Re,
  and a universal limiting state does not exist.
To decide between these alternative scenarios,
  Barenblatt and Goldenfeld computed best fits 
 of (\ref{asym}) to hot-wire data 
 from a large wind tunnel 
 and the atmosphere \cite{po}, 
  both for $p_0>0$ and $p_0=0$,
 and concluded that the data were 
 ``not inconsistent with either of the two 
 possibilities'' \cite{bg}.
In another attempt at deciding the matter,
 Sreenivasan \cite{sree} studied a large set of 
 hot-wire data, and concluded
 that the prefactor $p$ is ``{\it more or less\/}
 universal, {\it essentially\/}
  independent of the flow as well as the
  Reynolds number'' \cite{ital}.
 Nevertheless,
  he noted that the scatter in the data 
 was large, 
 and that to evince the behavior of $p$ at high Re
 one would have
 ``to cover a wide range of Reynolds numbers 
 in a single, well-controlled flow, and use instrumentation
 whose resolving power and quality remains 
 equally good in the entire range;'' unfortunately,
  ``such experiments and efforts are not yet in 
 the horizon at present'' \cite{sree}.
 To decide the matter, we intend to resort to
 experimental data on the friction coefficient 
 of rough conduits, $f$.
 These data  (e.g., \cite{niku}) 
 appear to be well suited to our purpose:
  they contain very little scatter
  and show beyond doubt that, for 
 any fixed wall roughness,
  the leading term of $f$ 
 is independent of Re at high Re.
  Further, the observed 
 dependence of the leading term of 
  $f$ on the roughness is described 
  accurately by a well-known empirical
  formula. The only problem, 
 to which we turn presently,
  is how to relate $f$ to 
 $\overline{(u_l)^2}$.

The friction coefficient of a conduit of 
 circular cross-section may be 
 defined as $f\equiv \tau/\rho V^2$, where $\tau$ is
 the shear stress on the wall of the 
conduit, $\rho$ the density of the liquid flowing
 through the conduit, and $V$ the average velocity
 of the flow. We seek to obtain an expression
 relating $f$ to the structure function
 of (\ref{asym}). Now (\ref{asym}) was 
 originally derived under the assumptions
 of isotropy and homogeneity, but 
 the turbulent flow in a conduit is both 
 anisotropic and inhomogeneous. 
 Nevertheless, recent research \cite{aniso}
 has established
 that (\ref{asym}) applies as well to flows that
 are neither isotropic nor homogeneous.
 Further, if $v_l$ denotes the characteristic 
 velocity of a turbulent eddy of 
 size $l$, we may identify 
 $v_l=\sqrt{\overline{(u_l)^2}}$, 
 where $\overline{(u_l)^2}$ is given by (\ref{asym})
 with $L=D$ (the diameter of the 
 conduit, which sets the largest lengthscale 
 in the flow)  (\cite{frisch}, chap.\ 7).
 Thus, it follows from (\ref{asym}) that, regardless of
 the value of $p_0$,  
 the smaller the eddy the lower its 
  velocity. With these considerations in mind,
 we now seek to derive an
 expression for $\tau$,
  the shear stress on the rough wall of the conduit.

\begin{figure}
\resizebox{2in}{!}{\includegraphics{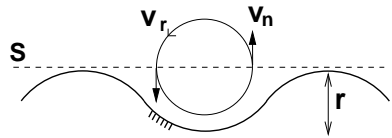}}
\caption{\label{fig1} 
  Immediate vicinity of a the rough-conduit wall with
  roughness elements of uniform size $r$ \cite{kd}. 
 The dashed line is the trace of the wetted surface 
 $S$. 
}
\end{figure}

Let us call $S$ the wetted surface 
 tangent to the peaks of 
 the roughness elements of the wall,
  Fig.~\ref{fig1}.  (We assume roughness 
 elements of uniform size $r$,
 as in Nikuradse's experiments \cite{niku}.) 
Then, for ${\rm Re}\gg 1$, the shear stress is effected by
 momentum transfer across $S$.
 Above $S$, the velocity of the flow scales
 with $V$, and the fluid carries
 a high horizontal momentum per unit volume
 ($\sim\rho V$). Below $S$, the velocity of the 
 flow is negligible, and the fluid carries
 a negligible horizontal momentum per unit volume.
 Now consider an eddy that straddles the wetted
 surface $S$
  (one half of the eddy is above $S$, the other half below).
 This eddy transfers fluid of high
 horizontal momentum downwards across $S$, 
 and fluid of negligible horizontal momentum
 upwards across $S$. 
 The net rate of momentum transfer across $S$ 
 is set by the velocity normal to $S$,
 which velocity is provided by the eddies that 
 straddle $S$. Therefore, if $v_n$ denotes 
 the velocity normal to $S$ provided
  by the {\it dominant\/} eddy that straddles
 $S$, then the shear stress effected by 
 momentum transfer across $S$ scales in the form
 $\tau\sim\rho\,V v_n$. Now the size 
 of the largest
 eddy that straddles $S$ scales with $r$, 
 the size of the roughness elements.
 This eddy provides a velocity 
 $v_r=\sqrt{\overline{(u_r)^2}}$
 normal to $S$,  where
   $\overline{(u_r)^2}$ is given by (\ref{asym}) 
 with $l=r$ and $L=D$. Smaller eddies 
 do provide a velocity normal to $S$,
 but these velocities are overwhelmed by the velocity 
 of the eddy of size $r$. 
 Thus,  $v_n\sim v_r$, 
 and the dominant eddy that straddles $S$
 is the largest eddy that straddles $S$. 
  We conclude that 
 $\tau\sim \rho\, v_r V$,
 and therefore $f\sim v_r/V=
\sqrt{\overline{(u_r)^2}}/V$.

To complete our derivation, we 
 relate $\varepsilon$ to $V$ and $D$ 
 using the phenomenological theory, which
 is based on two tenets 
 pertaining to the steady production
  of turbulent (kinetic) energy: 
 (1) The production occurs at the 
 lengthscale of the largest 
 eddies in the flow and (2) 
 The rate of production 
 is independent of the viscosity.
 From these tenets and the equality 
 of production and dissipation, it follows that 
 we can 
 obtain a scaling 
 expression for $\varepsilon$, 
 the rate of dissipation 
 of turbulent energy per unit mass of 
 liquid, in terms of 
 the velocity of the largest eddies
 (which $\sim V$) 
 and of the size of the largest eddies 
 (which $\sim D$) \cite{landau}.
 The largest eddies possess a kinetic
  energy per unit mass $e\sim V^2$
 and a turnover time $t\sim D/V$.
 These eddies persist for a time
 $t$, whereupon they split into 
  eddies of size $\sim D/2$,
  thereby transferring their energy
 to smaller lengthscales. For the steady state 
 to be preserved, a new set of large eddies 
  must be produced at time
 intervals $t$, implying that
  $\varepsilon= e/t\sim V^3/D$ \cite{upper}.
  By using $\varepsilon\sim V^3/D$, 
  $l=r$, and $L=D$ in (\ref{asym}), 
 and substituting the result in  
 $f\sim \sqrt{\overline{(u_r)^2}}/V$, we
  obtain
 \begin{equation} \label{ef}
 f \sim (r/D)^{1/3+\alpha/2} 
 \left(p_0 +{{p_1}\over{\ln{\rm Re}}}\right)^{1/2}, 
 \end{equation}
  an expression relating 
 $f$ to the parameters $p_0$, $p_1$,
 and $\alpha$ of the asymptotic 
 expansion of the structure function.

\begin{figure}
\resizebox{2.8in}{!}{\includegraphics{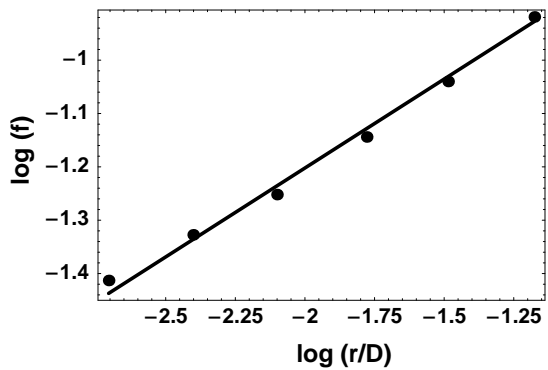}}
\caption{\label{fig2} A comparison of Nikuradse's data
 with Strickler's expression.
 The straight line is 
 $\log(f)=-0.54+\log(r/D)/3$. The data points correspond
  to the highest Re ($\approx 10^6$) tested
by Nikuradse. 
}
\end{figure}

 Now consider a conduit of fixed 
  roughness, $r/D= {\rm const} \ll 1$.
  If $p_0=0$ 
 the leading term of $f$ depends on Re: 
 $f_{\rm lt} \sim 
 \sqrt{p_1} \,(r/D)^{1/3+\alpha/2} (\ln{\rm Re})^{-1/2}$.
  Thus, for $p_0=0$ the friction coefficient
  vanishes asymptotically at high Re,
  a conclusion that is at odds with all
  experimental data on rough-conduit flows. 
 On the other hand, 
 if $p_0>0$ 
 the leading term of $f$ is independent of Re:
 $f_{\rm lt}\sim
  \sqrt{p_0}\, (r/D)^{1/3+\alpha/2}$.
 Thus, for $p_0>0$ the friction coefficient
  tends to a positive constant at high Re, 
  a conclusion
 that is qualitatively 
 consistent with all experimental data
 on rough-conduit flows.
 Further, in the case $\alpha=0$, 
 $f_{\rm lt}\sim
  \sqrt{p_0}\, (r/D)^{1/3}$,
  which we recognize as Strickler's
 empirical expression for the
  friction coefficient of a
   conduit of roughness $r/D$ 
 at high Re \cite{stri,bla}. 
 In the case $\alpha\neq 0$,
 $f_{\rm lt}\sim
  \sqrt{p_0}\, (r/D)^{1/3+\alpha/2}$,
 which is a generalized form of
 Strickler's empirical expression 
 that accounts for the effect of 
 intermittency. Given that the experimental
 data can be fitted very well even if 
 $\alpha$ is set to zero [Fig.~\ref{fig2}], 
 we infer that $1/3\gg |\alpha|/2$, or 
  $|\alpha|\ll 2/3$, consistent with 
 the available estimates 
   of $\alpha$ \cite{inter}.

From the previous paragraph, we conclude
 that equation (\ref{ef})
 together with the experimental data embodied
 by Strickler's empirical 
 expression allows us to establish
 that $p_0>0$. Therefore, 
  the leading term of the structure function
  $\overline{(u_l)^2}$ is independent of Re, and the 
  universal limiting state known as fully 
 developed turbulence exists
 at high Reynolds numbers. 

The logic of our reasoning so far
  has been the following. We have shown
 theoretically that in a rough conduit 
 $\lim_{{\rm Re}\rightarrow\infty} f=0$
  unless  $p_0>0$. In other words, we have 
 shown  that $p_0>0$
 is a {\it necessary\/} condition for 
 the friction 
 coefficient of a rough conduit
  to tend to a positive constant at high Re.
 Then, given the unequivocal experimental 
 evidence that in a rough conduit 
 $\lim_{{\rm Re}\rightarrow\infty} f={\rm const} >0$,
 we have concluded that it must be that $p_0>0$.
 Interestingly, $p_0>0$ is not in general 
 a {\it sufficient\/} condition for the friction 
 coefficient of a conduit to tend to a positive 
 constant at high Re: the conduit must be rough.
 To elucidate this statement, 
  we recall our scaling expression for the
  shear stress, 
  $\tau \sim \rho v_r V$. This expression indicates that 
 the momentum transfer is dominated by the 
 eddies of size $r$---the same size as the roughness 
 elements. Now in a rough-conduit flow
 of sufficiently high Re, $r$ exceeds the 
  viscous lengthscale, $r>\eta$. In fact,
 from $\eta\equiv\nu^{3/4}\varepsilon^{-1/4}$
 (where $\nu$ is the kinematic viscosity) and  
 $\varepsilon\sim V^3/D$ we can write
  $\eta/D\sim (\nu /VD)^{3/4} ={\rm Re}^{-3/4}$ 
 and conclude that for any given $r$ 
 the condition $r>\eta$ holds at sufficiently
 high Re. If Re is increased
 further, $\eta$ lessens and newer, smaller
 eddies populate the flow and become jumbled with
 the older eddies. Yet the momentum transfer
 continues to be dominated by eddies of 
 size $r$, and  $f$ 
 remains invariant. 
 This argument explains
 the behavior of $f$ for
  all rough conduits,
  no matter how small the
   the roughness elements. 
 If $r=0$, however, the condition $r>\eta$ cannot 
 be attained, even at extremely high Re.
 Thus in a smooth conduit  \cite{roughness}
 the momentum transfer will always be
 dominated by the smaller eddies in 
 the inertial range, whose size scales
 with $\eta$;
 since $\eta$ lessens as Re increases, 
 the momentum transfer will be dominated by ever 
 smaller eddies as Re increases, and 
  $\lim_{{\rm Re}\rightarrow\infty} f=0$---even though
 $p_0>0$, and the turbulence will be fully developed
 at high Re. To verify this conclusion mathematically,
  we study (\ref{ef}) for  
  $r\sim\eta$. By substituting 
   $\eta/D\sim {\rm Re}^{-3/4}$
 in place of $r/D$, we obtain 
 \begin{equation} \label{efr}
 f \sim {\rm Re}^{-1/4-3\alpha/8} 
 \left(p_0 +{{p_1}\over{\ln{\rm Re}}}\right)^{1/2}.
 \end{equation}
 In accord with our discussion
 above, (\ref{efr}) indicates that 
 in a smooth conduit the friction coefficient
  vanishes asymptotically at high Re,
 whether $p_0=0$ or $p_0>0$.
 Had we to decide between $p_0=0$ and
 $p_0>0$ on the basis of (\ref{efr}) and experimental data,
 the answer would not be clear-cut, because
 $(\ln{\rm Re})^{-1/2}$ varies but very slowly at high Re.
 Nevertheless, we have established previously
 that $p_0>0$. It follows that 
 $f_{\rm lt} \sim \sqrt{p_0}\,{\rm Re}^{-1/4-3\alpha/8}$,
  which in the case $\alpha=0$ 
  coincides with Blasius's
 empirical expression for the friction 
 coefficient of a 
 smooth conduit at high Re \cite{bla}.
 In the case $\alpha\neq 0$,
 $f_{\rm lt} \sim \sqrt{p_0}\,{\rm Re}^{-1/4-3\alpha/8}$
 is a generalized form of
 Blasius's empirical expression 
 that accounts for the effect of 
 intermittency. 

From the previous paragraph, we conclude
 that if $p_0>0$ (as we established before),
 our theoretical predictions 
 are in accord with the experimental data 
 embodied by Blasius's empirical expression.

In a recent monograph 
 \cite{tsinober}, 
 Tsinober has drawn a distinction
 between 
  ``{\it possibly\/} universal 
 properties of small scale turbulence'' and 
``quantitative universal properties
 of turbulence at high Re'' \cite{ital}.
 The former were first envisioned in the 
  early history of turbulence physics; 
 they are embodied by the 
 asymptotic scenarios of 
 Kolmog\'orov, K41 and K62, that define the
 possibly universal limiting state of 
 fully developed turbulence. 
The latter have long been known to
 engineers; they are manifest in a few 
  properties of turbulent flows that become independent of Re
 at high Re, including 
 the drag of bluff bodies and the
  friction coefficient of rough
 conduits, among others \cite{epstests}.
 Here 
 we have established a
 relation between the two \cite{newpaper}.
 In particular, we have found 
 that the existence of fully developed 
 turbulence is a 
 necessary condition for the 
 friction coefficient of rough conduits
 to tend to positive constants at high Re, as 
 seen 
 in experiments. In addition,
 we have found that the existence of fully developed 
 turbulence is compatible with 
  the experimental evidence that
  the friction coefficient of smooth conduits
 tends to zero at high Re.
 On the basis of our findings, fully developed 
 turbulence may be termed 
 a ``universal property of small scale 
 turbulence,'' without qualification.

\begin{acknowledgments}
 We thank J.\ W.\ Phillips for kindly
reading our manuscript and suggesting
ways of improving it.
\end{acknowledgments}


\begin{thebibliography}{}
\bibitem{orders}{A similar expression holds for the structure 
function of order $n$,
 $\overline{(u_l)^n} = P_n[{\rm Re},l/L]\,(\varepsilon l)^{n/3}$, 
 but we need concern ourselves only with the structure function 
 of order 2.}
\bibitem{fd}{Barenblatt and Goldenfeld wrote
 ``The mathematical expression of the assumption of
 a limiting state of fully developed turbulence
 is that statistical averages of the flow
 exhibit complete similarity with respect to Re'' \cite{bg}.}
\bibitem{natu}{For example, the form of the structure
 function should not change if we
 used the Taylor microscale Reynolds number, 
 ${\rm Re}_\lambda$, in lieu of a global Re.}
\bibitem{uni}{Note that if 
 $p_0>0$ and $p_1\neq 0$ we continue to say 
 that fully developed
 turbulence exists, even though the structure function 
 depends on Re, and the state of fully developed
 turbulence can be attained
 only asymptotically. This scenario has been termed
 ``asymptotic Kolmog\'orov scaling.'' Given that K41
  was predicated on the asymptotic
  condition ${\rm Re}\gg 1$, we believe that
  in  its original formulation
  K41 was not meant to imply anything stronger
 than asymptotic Kolmog\'orov scaling. 
 Hence, we characterize K41 and K62 as
 ``asymptotic scenarios.''}
\end{thebibliography}

\end{document}